\documentclass[twocolumn,showpacs,preprintnumbers,amsmath,amssymb,superscriptaddress,prb,floatfix]{revtex4}

\usepackage{graphicx}
\usepackage{bm}

\begin{document}

\title{Resonant elastic soft x-ray scattering in oxygen-ordered YBa$_2$Cu$_3$O$_{6+\delta}$}

\author{D. G. Hawthorn}
\affiliation{Department of Physics and Astronomy, University of Waterloo, Waterloo, N2L 3G1, Canada}

\author{K. M. Shen}
\affiliation{Laboratory of Atomic and Solid State Physics, Department of Physics, Cornell University, Ithaca NY 14853, USA}

\author{J. Geck}
\affiliation{Leibniz Institute for Solid State and Materials Research IFW Dresden, Helmholtzstrasse 20, 01069 Dresden, Germany}

\author{D. C. Peets}
\altaffiliation{Present address: Max Planck Institute for Solid State Research, D-70569 Stuttgart, Germany}
\affiliation{Department of Physics and Astronomy, University of British Columbia, Vancouver,V6T 1Z4, Canada}

\author{H. Wadati}
\altaffiliation{Present address:  Department of Applied Physics and Quantum-Phase Electronics Center (QPEC), University of Tokyo, Hongo, Tokyo 113-8656, Japan}
\affiliation{Department of Physics and Astronomy, University of British Columbia, Vancouver,V6T 1Z4, Canada}

\author{J. Okamoto}
\affiliation{National Synchrotron Radiation Research Center, Hsinchu 30076, Taiwan}

\author{S.-W. Huang}
\affiliation{National Synchrotron Radiation Research Center, Hsinchu 30076, Taiwan}

\author{D. J. Huang}
\affiliation{National Synchrotron Radiation Research Center, Hsinchu 30076, Taiwan}

\author{H.-J. Lin}
\affiliation{National Synchrotron Radiation Research Center, Hsinchu 30076, Taiwan}

\author{J. D. Denlinger}
\affiliation{Advanced Light Source, Lawrence Berkeley National Laboratory, Berkeley, CA 94720-8229, USA}

\author{Ruixing Liang}
\affiliation{Department of Physics and Astronomy, University of British Columbia, Vancouver,V6T 1Z4, Canada}

\author{D. A. Bonn}
\affiliation{Department of Physics and Astronomy, University of British Columbia, Vancouver,V6T 1Z4, Canada}
 
\author{W. N. Hardy}
\affiliation{Department of Physics and Astronomy, University of British Columbia, Vancouver,V6T 1Z4, Canada}

\author{G. A. Sawatzky}
\affiliation{Department of Physics and Astronomy, University of British Columbia, Vancouver,V6T 1Z4, Canada}

\date{\today}

\begin{abstract}
Static charge-density wave (CDW) and spin-density wave (SDW) order has been convincingly observed in La-based cuprates for some time.  However, more recently it has been suggested by quantum oscillation, transport and thermodynamic measurements that density wave order is generic to underdoped cuprates and plays a significant role in YBa$_2$Cu$_3$O$_{6+\delta}$ (YBCO).  We use resonant soft x-ray scattering at the Cu $L$ and O $K$ edges to search for evidence of density wave order in Ortho-II and Ortho-VIII oxygen-ordered YBCO.     We report a null result -- no evidence for static CDW order -- in both Ortho-II and Ortho-VIII ordered YBCO.  While this does not rule out static CDW order in the CuO$_2$ planes of YBCO, these measurements place limits on the parameter space (temperature, magnetic field, scattering vector) in which static CDW order may exist.  In addition, we present a detailed analysis of the energy and polarization dependence of the Ortho-II superstructure Bragg reflection [0.5~0~0] at the Cu $L$ edge.  The intensity of this peak, which is due to the valence modulations of Cu in the chain layer, is compared with calculations using atomic scattering form factors deduced from x-ray absorption measurements.   The calculated energy and polarization dependence of the scattering intensity is shown to agree very well with the measurement, validating the approach and providing a framework for analyzing future resonant soft x-ray scattering measurements.

\end{abstract}




\maketitle

In the cuprate superconductors, it has long been recognized that incommensurate spin density wave (SDW) and charge density wave (CDW) order co-exists or competes with the superconducting phase.\cite{Vojta09}  This CDW/SDW order is most clearly manifested in 1/8-doped La-based cuprates where CDW and SDW are stabilized and made static by a low temperature tetragonal (LTT) lattice distortion.\cite{Tranquada95,Zimmermann98,Kimura03}  In other cuprates, checkerboard-like static density wave order, different from the stripe ordering in La-based cuprates, has been observed with surface-sensitive scanning tunnelling microscopy measurements.\cite{Kohsaka08,Hanaguri04}  More recently, static SDW order has also been observed in low-doped (below $p=1/8$) YBa$_2$Cu$_3$O$_{6+\delta}$ (YBCO) by neutron scattering at low temperatures and in high magnetic fields.\cite{Hinkov08,Haug09}  In addition to these direct observations of density wave order, other indirect measurements have suggested that density wave order is more generic to the cuprates than was previously believed.   These include recent quantum oscillation measurements detecting the presence of unexplained electron pockets in underdoped YBCO\cite{Doiron07,Bangura08,Yelland08,Jaudet08,Sebastian08} that may result from density wave order causing a Fermi surface reconstruction,\cite{Millis07} a striking similarity in the Hall-coefficient between YBCO and stripe-ordered LSCO,\cite{LeBoeuf07} and anisotropy in the Nernst co-efficient suggestive of unidirectional order.\cite{Daou09}   Despite this evidence, it is not yet apparent how generic static CDW and/or SDW order is to the cuprates and ultimately what role these density wave orders play in the superconductivity.\cite{Emery97,Kivelson98,Berg09}  Furthermore, it is still open to debate whether disorder, structural distortions and magnetic field, which seem to stabilize density wave order in La-based cuprates, are critical to making the density wave order static in other cuprates.\cite{Tranquada95,Hucker10,Lake02}

Some of these issues would be resolved by an observation of static CDW order in YBCO, which is structurally different and generally less disordered than LSCO.  In YBCO, an important candidate to search for CDW order is Ortho-VIII ordered YBCO.  In Ortho-VIII ordered YBCO, oxygen atoms are ordered in a 11011010 pattern in the oxygen chain layer  (1 and 0 indicate the presence or absence of intercalated oxygen in a given chain).\cite{Beyers89,Zimmermann03}  This oxygen doping level not only has a hole doping in the CuO$_2$ planes near $p$ = 1/8, the doping where stripe ordering is stabilized in La-based cuprates, it also has a natural periodicity provided by the oxygen ordering that may be commensurate with CDW/SDW ordering.  Like the LTT structural phase in the La-based cuprates, this ordering may form a structural template that stabilizes static stripe ordering.  This same material also exhibits a large change in the Hall coefficient that may be associated with density wave order.\cite{LeBoeuf07,Chang10}  

\begin{figure}[t]
\centering
\resizebox{2.25in}{!}{\includegraphics{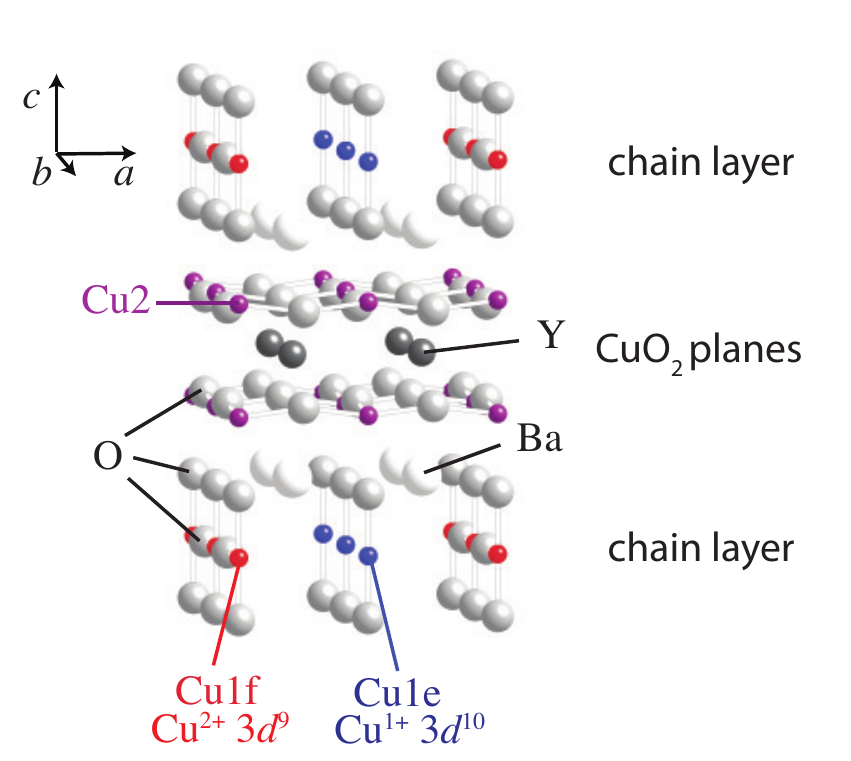}}
\caption{(color online)The crystal structure of Ortho-II ordered YBa$_2$Cu$_3$O$_{6.5}$.  YBCO has Cu sites in both the CuO$_2$ planes (Cu2) and the chain layer (Cu1).  In Ortho-II ordered YBCO the dopant oxygen atoms in the chain layer are ordered forming alternating rows of full and empty chains.   This induces an alternation in the valence of Cu1 from Cu$^{2+}$ in the full chains (Cu1f, red) to Cu$^{1+}$ in the empty chains (Cu1e, blue).}
\label{fig1_structure}
\end{figure}

Ortho-II ordered YBCO (see Fig.~\ref{fig1_structure}) presents another important candidate for CDW ordering.  In this material, electron pockets have been observed that may indicate a density-wave-order induced Fermi-surface reconstruction.\cite{Doiron07,Jaudet08,Sebastian08}  In addition, there are several reports from diffraction measurements of a coupling between the planes and chains where oxygen order in the chains induces a CDW in the planes.\cite{Islam02,Feng04}  However, these latter results remain controversial, with one result\cite{Islam02} having failed to be reproduced in subsequent measurements.\cite{Strempfer2004}

Resonant soft x-ray scattering is an ideal probe to search for CDW in YBCO.  Recently resonant soft x-ray scattering (RSXS) has emerged as a powerful probe of spin, charge and orbital order in a variety of transition metal oxide materials.\cite{Abbamonte04,Abbamonte05,Wilkins03,Thomas04,Schusler05,Huang06}  By tuning the x-ray energy to an x-ray absorption edge, the atomic scattering form factor, $f(\omega)$, is enhanced, producing a scattering measurement that is tunably sensitive to specific atomic orbitals (e.g.. O $2p$ or Cu $3d$ states) as well as the spin and orbital symmetry of those orbitals (e.g. the Cu $3d_{x^2-y^2}$ states).  This sensitivity can be applied to examine subtle ordering phenomena, such as charge stripes in cuprate superconductors,\cite{Tranquada95,Abbamonte05,Fink09} by tuning to the x-ray absorption edges that probe the most relevant orbitals in the material.  In the cuprates, these edges are the O $K$ edge and the Cu $L$ edges, which probe the unoccupied O 2$p$ and Cu 3$d$ states respectively -- the states near $E_F$ that are most responsible for the the low energy physics of the cuprates, including HTSC and density-wave order.  RSXS has already been used to explore CDW order in La-based cuprates doped with Ba or Eu, arguably providing the most direct measurements yet of static CDW order in the cuprates.\cite{Abbamonte05, Fink09}  In other cuprates, however, there has not been any conclusive evidence from diffraction experiments of static charge ordering in the bulk.  In Ca$_{2-x}$Na$_x$CuO$_2$Cl$_2$ a null result was reported by Smadici et al.\cite{Smadici07A} in their search using RSXS for checkerboard order that has been observed by STM.\cite{Hanaguri04}

In this manuscript we present our search for CDW in Ortho-VIII and Ortho-II ordered YBCO using resonant elastic soft x-ray scattering.  In both phases we report a null result -- no evidence for CDW ordering in the CuO$_2$ planes.  This result restricts, but does not rule out, scenarios for static CDW order in YBCO.   This null result runs contrary to the conclusions reached by Feng et al., who claim a substantial coupling between the CuO$_2$ planes and chain layer from the polarization dependence of RSXS measurements.\cite{Feng04}  Although our measurements are similar to Feng et al., a more detailed analysis of the polarization and energy dependence of the scattering presented here indicates that charge ordering in the CuO$_2$ planes cannot be resolved from either measurement.

This detailed analysis of the RSXS energy lineshape also serves as an important test case for analyzing resonant soft x-ray scattering lineshapes.  From the energy dependence of RSXS one can distinguish between and distill microscopic details of magnetic order, orbital order, charge order and lattice distortions.  Although there is a theoretical basis for understanding the spectroscopy of resonant soft x-ray scattering, there are as yet few rigorous tests of the technique that clearly reproduce the polarization and energy dependence of the scattering intensity.  In some instances, configuration-interaction calculations have reproduced resonant scattering spectra.\cite{Schusler05, Haverkort08}  In other instances, such as La$_{2-x}$Ba$_x$CuO$_4$ or La$_{1.8-x}$Eu$_{0.2}$Sr$_x$CuO$_4$, XAS data from related materials has been used to determine the energy dependence of the atomic scattering form factor and subsequently calculate the RSXS.\cite{Abbamonte05,Fink09} However, using this latter approach the agreement between experiment and simple calculations is poorer than one would anticipate in RSXS on La$_{2-x}$Ba$_x$CuO$_4$ or La$_{1.8-x}$Eu$_{0.2}$Sr$_x$CuO$_4$, particularly given the relative simplicity of the electronic structure of the cuprates.  Here we employ a similar approach to interpreting the resonant scattering lineshape by investigating the [0.5~0~0] Ortho-II superstructure peak of YBa$_2$Cu$_3$O$_{6.5}$, which has non-trivial energy and polarization dependence.  In this case the Ortho-II oxygen ordering results in a valence modulation of Cu in the chain layer that by symmetry does not induce a corresponding displacement of Cu away from the the ideal positions in the unit cell.  Furthermore, the atomic scattering form factor for Cu in the chain layer can be determined using XAS measurements on YBCO with either entirely full chains ($\delta = 1$) or entirely empty chains ($\delta = 0$).  We find excellent agreement between experiment and model calculations based on the $f(\omega)$ deduced from XAS measurements.

\section{Experimental details}

High-quality YBCO single crystals were grown in BaZrO$_3$ crucibles by a copper-rich self-flux technique,\cite{Liang98} then annealed under carefully-controlled oxygen partial pressures to achieve the desired oxygen contents and ordering.\cite{Liang00}   
X-ray diffraction measurements were performed on the as-grown $\langle 100 \rangle$ face of  Ortho-II ordered YBa$_2$Cu$_3$O$_{6.5}$  with linearly polarized light having incident polarization parallel or perpendicular to the scattering plane, with $\vec{\epsilon}\perp \vec{c}$ or $\vec{\epsilon}\parallel \vec{c}$ respectively.  In the following the orthorhombic unit cell of oxygen disordered YBCO is used with $a$ = 3.831 \AA, $b$ = 3.887 \AA ~and $c$ = 11.75 \AA.
Four single-crystal samples of Ortho-VIII ordered YBa$_2$Cu$_3$O$_{6.63}$ were also measured.  The samples were prepared with different orientations (two samples with an $ab$ face and two with a $bc$ face) in order to explore as much of reciprocal space as possible.  All measurements were performed in a pressure of $\lesssim1\times10^{-8}$ Torr and the Ortho-II sample was measured at a temperature of $\sim78$ K.   All resonant x-ray scattering measurements were performed at the 05B3 beamline of the National Synchrotron Radiation Research Center in Taiwan.

X-ray absorption (XAS) measurements of YBa$_2$Cu$_3$O$_{6+\delta}$ with $\delta$ = 0, 0.5 and 1 were performed on twin-free samples with the incident light oriented normal to the sample surface.   For each doping three samples were measured with the orientation of the linear polarization aligned variously along the three crystallographic axes ($\vec{\epsilon} \parallel \vec{a}, \vec{b}$ and $\vec{c}$).  The samples were polished with 0.05 $\mu $m alumina grit and etched in Br diluted in anhydrous ethanol prior to measurement to clean the surface.  The measurements were performed using total fluorescence yield (TFY) at beamline 8.0.1 \cite{Jia95} of the Advanced Light Source at room temperature and in pressure of $<1\times10^{-8}$ Torr.

\section{XAS in YBCO}
\begin{figure}[th]
\centering
\resizebox{\columnwidth}{!}{\includegraphics{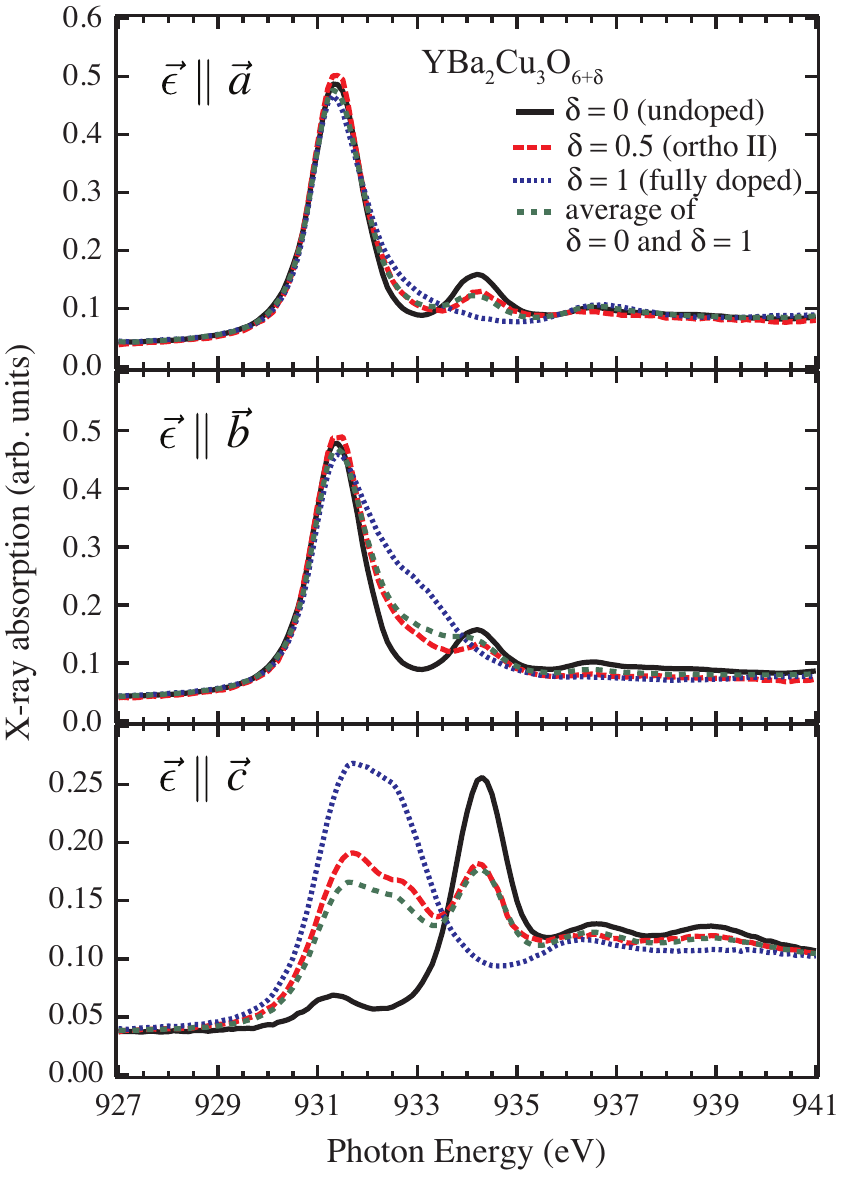}}
\caption{(colour online) The x-ray absorption (total fluorescence yield) of YBa$_2$Cu$_3$O$_{6+\delta}$ at the Cu $L_{3}$ absorption edge for $\vec{\epsilon} \parallel \vec{a}$, $\vec{\epsilon} \parallel \vec{b}$ and $\vec{\epsilon} \parallel \vec{c}$.  Three samples are shown with oxygen doping $\delta = 0$ (black), 0.5 (red) and 1 (blue).  The average of the XAS of the undoped ($\delta = 0$) and fully doped ($\delta = 1$) samples shown in green agrees well with the measured spectra of Ortho-II ordered YBCO ($\delta = 0.5$), supporting the premise that Ortho-II ordered YBCO consists of alternating full and empty chains.}
\label{fig2}
\end{figure}

In YBa$_2$Cu$_3$O$_{6+\delta}$, hole doping into the CuO$_2$ planes can be achieved by intercalating oxygen into the chain layer (see Fig.~\ref{fig1_structure}).  The electronegativity of the intercalated O$^{2-}$ removes electrons from Cu in the chain layer (Cu1) changing the formal valence of Cu1 from Cu$^{1+}$ to Cu$^{2+}$.  The oxygen doping also dopes holes into the CuO$_2$ planes.  XAS is sensitive to this change in valence and also to the orbital symmetry of the doped holes. 

In Fig.~\ref{fig2}, XAS at the Cu $L_3$ edge is shown for samples with oxygen dopings of $\delta = 0$ (empty chains), $\delta = 0.5$ (Ortho-II) and $\delta \simeq 1$ (full chains) for $\vec{\epsilon} \parallel \vec{a}, \vec{b}$ and $\vec{c}$.  Our XAS results are similar to previous XAS measurements of YBCO.\cite{Nucker95,Fink94,Merz98}  

For $\delta = 0$, the Cu2 in the CuO$_2$ planes are in a 3$d^9$ ground state configuration, having a single hole in the Cu 3$d_{x^2-y^2}$ orbital.  This is evidenced by a large peak in the XAS at 931.3 eV for $\vec{\epsilon} \parallel \vec{a}/ \vec{b}$ that is significantly weaker for $\vec{\epsilon} \parallel \vec{c}$ (a small contribution is present for $\vec{\epsilon} \parallel \vec{c}$ owing to some hybridization between states with 3$d_{x^2-y^2}$ and 3$d_{3z^2-r^2}$ symmetry).  In the CuO$_2$ chains the Cu1 is formally in a 3$d^{10}$ ground state, with a full $d$ shell.  However, a sharp peak is still observed in the XAS at 934.3 eV that is attributed to monovalent Cu, similar to the XAS in Cu$_2$O.\cite{Grioni92}

Doping additional oxygen into YBCO produces significant changes in the XAS.  As the oxygen content is changed from $\delta = 0$ to $\delta = 1$, additional holes are doped into the chain layer, converting Cu$^{+1}$ to Cu$^{+2}$ .  For $\delta = 1$ Cu1 has a 3$d^9$ ground state configuration with a hole in the state with 3$d_{y^2-z^2}$ symmetry and additional holes in a 3$d^9\underline{L}$ state, also with 3$d_{y^2-z^2}$ symmetry, similar to the $d^9$ +  $d^9\underline{L}$ state of the CuO$_2$ planes in hole doped cuprates.\cite{Chen92,Chen91,Fink94}  This configuration is evidenced by a pair of peaks at 931.7 eV ($d^9$) and 932.7 eV ($d^9\underline{L}$) that appear most clearly with $\vec{\epsilon} \parallel \vec{c}$, and also as a shoulder in the $\vec{\epsilon} \parallel \vec{b}$ spectrum.

In addition to the large changes corresponding to the doping of holes into the chain layer, oxygen doping also dopes holes into the CuO$_2$ planes.  As in other cuprates such as LSCO, \cite{Chen92,Pellegrin93} this doping produces a shoulder in the $\vec{\epsilon} \parallel \vec{a}$ or $\vec{\epsilon} \parallel \vec{b}$ spectrum (at 932.8 eV) corresponding to $d^{9}\underline{L}$.

In carefully annealed high-purity samples, the oxygen can be ordered into arrays of full and empty chains, producing alternating rows of CuO$_3$ (with Cu$^{2+}$, denoted as Cu1f) and CuO$_2$ (with Cu$^{1+}$, denoted as Cu1e).   The Ortho-II phase corresponds to alternating full and empty chains, as depicted in Fig.~\ref{fig1_structure}, whereas the Ortho-VIII phase corresponds to a repeating 11011010 ordering of full (1) and empty (0) chains.  For alternating full and empty chains, it is reasonable to expect the XAS of Ortho-II YBCO to be the average of the XAS of $\delta = 0$, with entirely empty chains, and $\delta = 1$, with entirely full chains.  As shown in Fig.~\ref{fig2}, this is approximately true; the average of $\delta = 0$ and 1 agrees well with the Ortho-II XAS.  This provides a reasonable basis to use the spectrum for $\delta = 0$ to determine $f_{Cu1e}$ and the spectrum for $\delta = 1$ to determine $f_{Cu1f}$.

\begin{figure}[th]
\centering
\resizebox{2.7in}{!}{\includegraphics{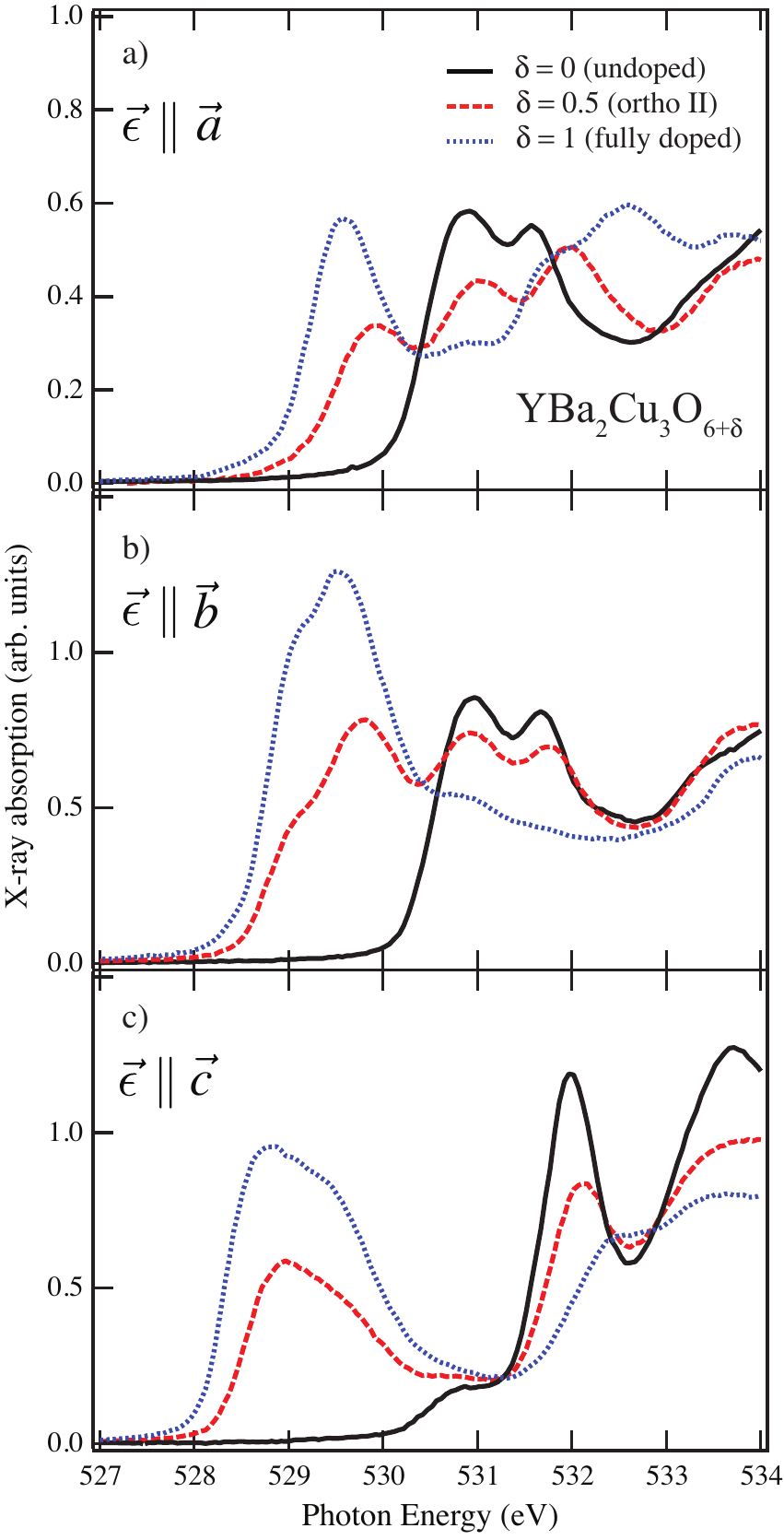}}
\caption{(colour online) Oxygen $K$ edge x-ray absorption for undoped ($\delta = 0$), fully doped ($\delta = 1$) and Ortho-II ordered YBCO ($\delta = 0.5$) measured by total fluorescence yield for $\vec{\epsilon} \parallel \vec{a}$,  $\vec{b}$ and $\vec{c}$.  For $\vec{\epsilon} \parallel \vec{a}$, the XAS pre-edge probes the unoccupied $p_x$ states that are hybridized with the Cu 3$d_{x^2-y^2}$ states of the CuO$_2$ planes.  For $\vec{\epsilon} \parallel \vec{c}$, the XAS pre-edge probes th unoccupied $p_z$ states  that are hybridized with the Cu 3$d_{y^2-z^2}$ states of the chain layer.   Measurements with $\vec{\epsilon} \parallel \vec{b}$ probe both the CuO$_2$ planes and the chain layer.  The peak around 529.5 eV for $\vec{\epsilon} \parallel \vec{a}$ is equivalent to the peak where the stripe scattering intensity is enhanced in RSXS measurements of La-based cuprates.\cite{Abbamonte05,Fink09}}
\label{figOKXAS}
\end{figure}

In Fig.~\ref{figOKXAS} XAS at the O $K$ edge is shown for fully doped, Ortho-II ordered, and undoped YBCO.   Results are similar to previous XAS measurements.\cite{Nucker95}  With $\vec{\epsilon} \parallel \vec{a}$ or $\vec{b}$ the pre-edge structure at 529.5 eV corresponds to O $p_x$ and $p_y$ states that are hybridized with the Cu $3d_{x^2-y^2}$ in the CuO$_2$ planes.  This structure is associated with the doped holes and is where resonant scattering is enhanced in investigations of stripes in La-based cuprates.\cite{Abbamonte05,Fink09}  For $\vec{\epsilon} \parallel \vec{c}$ ($\vec{b}$), the XAS has peaks associated with $3d_{y^2-z^2}$ states of the full chains at 528.8 eV (529.3 eV) and 529 eV (529.6 eV) for $\delta = 1$ and $\delta = 0.5$ respectively.  

\section{Search for CDW ordering in Ortho-VIII ordered YBCO}

In searching for CDW order in Ortho-VIII ordered YBCO a number of guidelines were considered:   

Firstly, the likely position in reciprocal space to observe CDW order is near [0.25, 0, $L$], where charge ordering peaks are observed in La-based cuprates.  However, the difference in structure between YBCO and La-based cuprates may lead to different structure factor for CDW order in the two materials.  CDW order is likely to be weakly correlated along the $c$-axis in YBCO, similar to La-based cuprates, leading to a broad structure in $L$.  This is especially true due to the weak correlations of the Ortho-VIII oxygen order along the $c$-axis.\cite{Zimmermann03}  However, the bilayer structure of YBCO may lead to CDW order that is strongly coupled within a bilayer but weakly coupled between neighbouring bilayers.  In this instance, scattering would remain broad in $L$ but may have a minimum at $L$ = 0 if the CDW is anti-correlated between the two planes of an individual Cu$_2$O$_4$ bilayer.  

Secondly, CDW order is likely only static at low temperature.   Based on Hall effect measurements, in which a downturn in the Hall coefficient has been argued to be a signature of the onset of stripe ordering, this ordering temperature is estimated to be $\simeq$ 70 K.\cite{LeBoeuf07}  Since the oxygen ordering temperature (430 K)\cite{Zimmermann03} is well above the expected CDW ordering temperature, the temperature dependence of any features can be used to attribute them to either CDW order or oxygen ordering.
\begin{figure}[t]
\centering
\resizebox{\columnwidth}{!}{\includegraphics{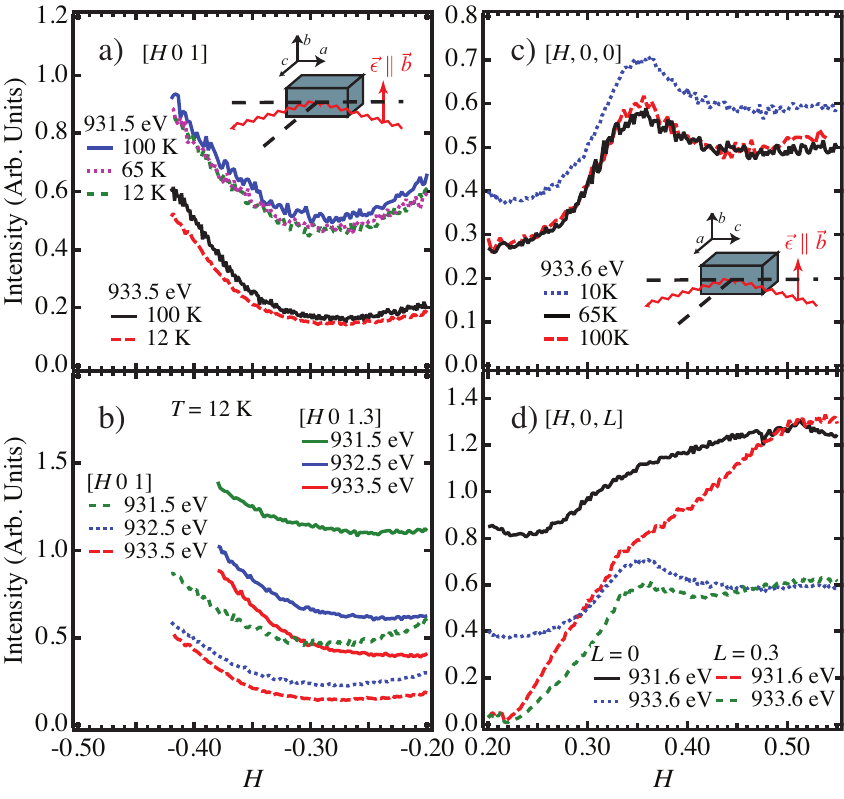}}
\caption{(colour online) Cu $L$ edge searches for stripe ordering in Ortho-VIII YBCO.  a,b)  The measured intensity through [$H$ 01] and [$H$ 0 1.3] with $\vec{\epsilon} \parallel b$ at several temperatures for different photon energies.  c,d)  The measured intensity through [$H$ 0 0] and [$H$ 0 0.3] with $\vec{\epsilon} \parallel b$ at several temperatures for different photon energies.   At 931.6 eV, the atomic scattering form factor for Cu2 in the CuO$_2$ planes should be enhanced, whereas for 933.6 eV, the enhancement is more likely associated with the Cu1 site in the chain layer.  In all cases, no superlattice peak is observed near $H$ = 1/4.  The sample geometry is depicted in the insets.}
\label{fig1_OrthoVIII}
\end{figure}

Lastly, as a function of photon energy and light polarization, the scattering intensity from CDW order is expected to be peaked at the peak in the Cu $L$ edge XAS (931.4 eV) and the pre-edge of the O $K$ edge (529.8 eV) XAS, as has been observed for stripes in LBCO\cite{Abbamonte05} and Eu-LSCO\cite{Fink09}.  In addition, since the CDW order is expected to occur in the CuO$_2$ planes where the doped holes occupy orbitals with $d_{x^2-y^2}$ symmetry, CDW order is most likely to be detected with the x-rays having a component of the light polarization in the $ab$ plane.  In YBCO, which has copper in both the CuO$_2$ planes and the chain layer, CDW in the planes and oxygen order in the chains can be distinguished by the energy and polarization dependence of the scattering intensity.

In Fig.~\ref{fig1_OrthoVIII} a) and b) we show RSXS scans in Ortho VIII YBCO through [$H$, 0, 1]  and [$H$, 0, 1.3] with $\vec{\epsilon} \parallel \vec{b}$ and at several photon energies through the Cu $L_3$ absorption edge.  No peak is observed near $H = -0.25$.  The observed variation in the background as a function of $H$ can be attributed to the x-ray fluorescence, which varies as a function of the incident photon energy and the angles of incidence and emission.\cite{Eisebitt93}

\begin{figure}[t]
\centering
\resizebox{2.5in}{!}{\includegraphics{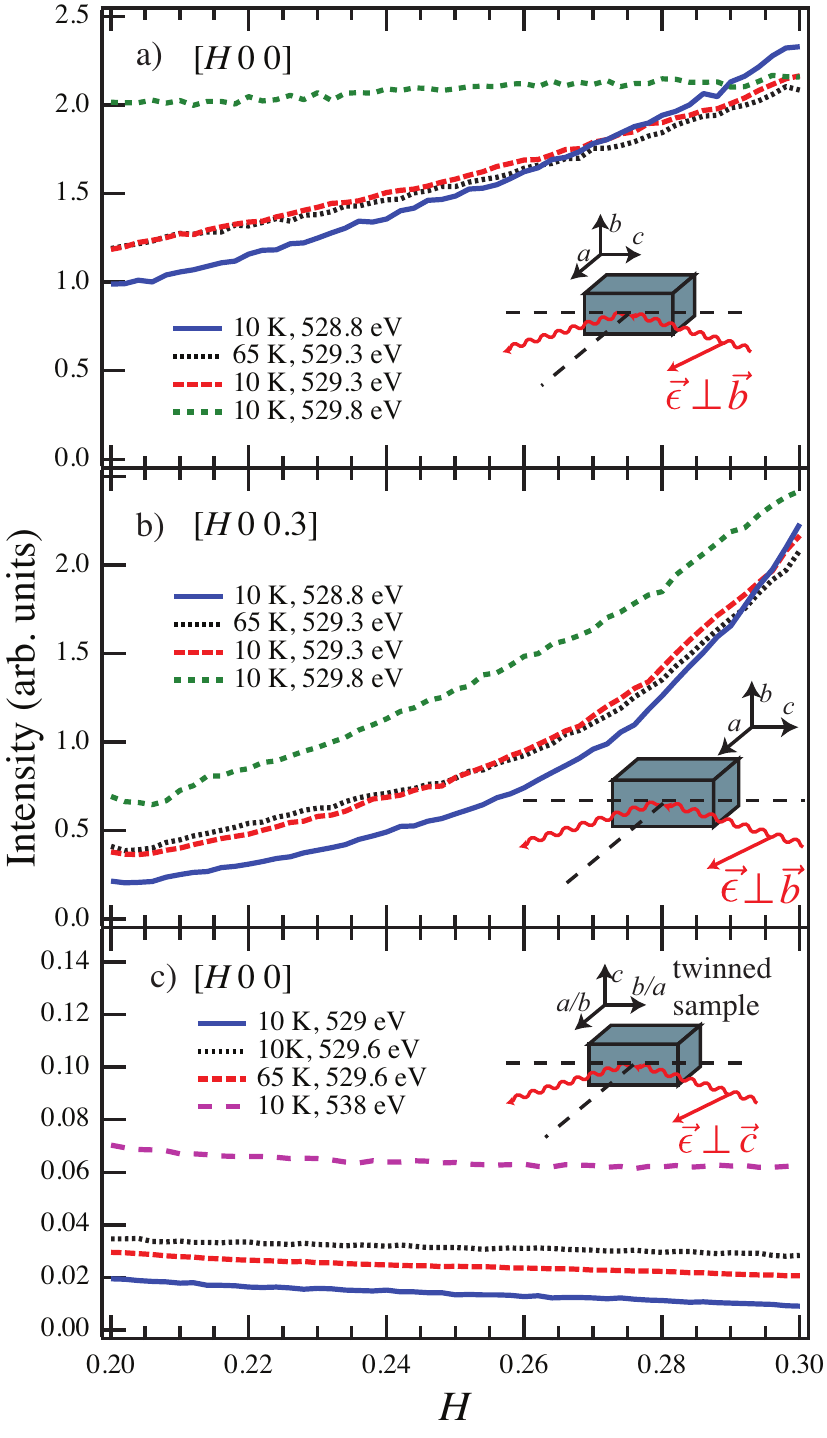}}
\caption{(colour online) Oxygen $K$ edge searches for CDW order in Ortho-VIII YBCO.  a-c)  The measured intensity through [$H$ 0 0] and [$H$ 0 0.3].  Sample geometry is depicted in the insets.  No superlattice peak is observed near $H$ = 1/4.  The sloping background in a) and b) is attributed to angle dependent x-ray fluorescence.}
\label{OrthoVIIIfig2-OK}
\end{figure}

In Fig.~\ref{fig1_OrthoVIII} c) and d) we show RSXS scans in Ortho VIII YBCO through [$H$, 0, 0] and [$H$, 0, 0.3] with $\vec{\epsilon} \parallel \vec{b}$ and at several photon energies through the Cu $L_3$ absorption edge.  At 933.6 eV, a peak is observed centred around $H$ = 0.365.  This $H$ position roughly corresponds to the $H = 3/8$ peak expected for the Ortho-VIII oxygen ordering.  For the Ortho-VIII order, the scattering intensity is peaked at $H$ = 3/8 and 5/8, with smaller peaks at $H$ = 1/2, 1/8 and 7/8.\cite{Zimmermann03}  This assignment is clarified by the energy and temperature dependence of this peak.  The peak is smaller at 931.6 eV corresponding to the states in the CuO$_2$ planes but predominant at 933.6 eV, corresponding to states in the chain layer.  As discussed below, the detailed energy dependence of the scattering intensity through the $L_3$ and $L_2$ absorption edges corresponds well with the calculated scattering intensity for the Ortho-VIII oxygen ordering in the chain layer, confirming its origin.  This peak exhibits weak temperature dependence.  However, the sample was small relative to the beam spot and small motions of the sample may have led to temperature dependent changes in the detected intensity.  Normalizing the peak to the background indicates no intrinsic temperature dependence.  Additional upturns in the measured intensity are also observed around $H = 0.2$.  By moving the sample position relative to the beam, these features were identified as a specular reflection from the irregularly shaped edge of the sample.

In Fig.~\ref{OrthoVIIIfig2-OK} we show RSXS scans through [$H$ 0 0] and [$H$ 0 0.3] with $\vec{\epsilon} \perp \vec{b}$ through the O $K$ edge. Similar to the measurements at the Cu $L$ edge, no superlattice peak is observed on the O $K$ edge at $H$ = 0.25.  A sloping background is observed in Figs.~\ref{OrthoVIIIfig2-OK} a) and b).  Like the Cu $L$ edge, this can be attributed to the change in x-ray fluorescence as a function of energy, angle of incidence and angle of emission.

While this null result does not exclude the possibility of static CDW order in YBCO, it places restrictions on the possibilities if static CDW order is indeed present.  If static CDW order is present, it should have a magnitude that is too weak in intensity and broad in $Q$ to be observed above the noise in the present measurements.  A simple and somewhat naive attempt to place a quantitative restriction on the magnitude of stripe ordering can be made by comparing to the intensity of the CDW superlattice peak observed in LBCO using RSXS\cite{Abbamonte05} where the CDW superlattice peak measured at the peak of the Cu $L_3$ edge has a maximum intensity of 175 counts/s on a fluorescent background of 450 counts/s, giving a peak/background of 39\%.  In our experiment, we would expect to observe a peak that appears above our noise level, which is $\sim 1.5\%$ of the fluorescent background.  All else being equal, this comparison indicates that, if present, static CDW order in YBCO would have a magnitude $\sim25 \times$ weaker than 1/8 doped LBCO.  

If static CDW order is not present, a likely possibility is that CDW order in YBCO is dynamic rather than static.\cite{Kivelson98,Kivelson03}  YBCO may be similar to La$_{2-x}$Sr$_x$CuO$_4$ (as opposed to Nd-doped LSCO, LBCO or Eu-doped LSCO) where SDW order\cite{Suzuki98,Kimura99,Kimura00,Fujita02,Chang08,Wakimoto99,Matsuda00} has been observed, but the static CDW order that should accompany SDW order has not.\cite{Fujita02b,Kimura03,Vojta09}  As noted by Kimura et al. \cite{Kimura03}, this can possibly be reconciled by considering that charge stripe order is intrinsically disordered, rendering it too weak to be detected by x-ray and neutron diffraction measurements.  Unfortunately, static SDW order cannot be probed with RSXS due to restrictions in the range of reciprocal space accessible at the low photon energies of the O $K$ or Cu $L$ edges (in LSCO SDW peaks are observed split off ($\pi$, $\pi$, $L$), which is too large in $Q$ for the present measurements.)  It may also be that an applied magnetic field is necessary to stabilize density wave order in YBCO.  In La$_{2-x}$Sr$_x$CuO$_4$ an applied magnetic field enhances SDW order\cite{Lake02,Khaykovich03,Khaykovich05,Chang08} and the Hall effect, Nernst and quantum oscillation measurements that suggest density wave order in YBCO are performed in an applied magnetic field (although the downturn in the Hall co-efficient seems to persist to zero-field over at least a finite range of temperature).\cite{LeBoeuf07}

\begin{figure}[th]
\centering
\resizebox{\columnwidth}{!}{\includegraphics{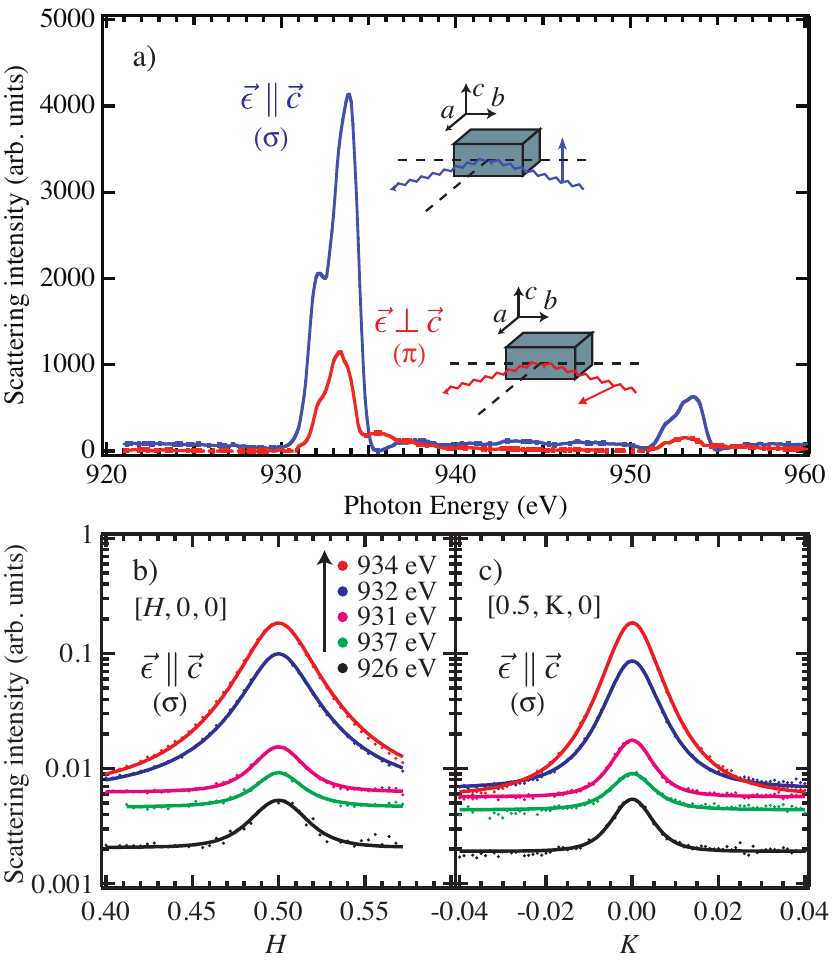}}
\caption{(colour online) Resonant x-ray scattering from the Ortho-II oxygen superstructure of YBa$_2$Cu$_3$O$_{6.5}$ at the Cu $L_{2,3}$ absorption edges.  a)  The intensity of the [0.5~0~0] superstructure peak vs. incident photon energy for $\sigma$ (blue) and $\pi$ (red) incident polarization.  b) and c) The scattering intensity through [0.5 0 0] vs. $H$ and $K$, respectively.  The lines are the result of fitting the data to equation~\ref{eq:fit}.}
\label{fig1}
\end{figure}
\section{RSXS lineshape analysis of oxygen superlattice peaks in oxygen ordered YBCO}
RSXS was also performed on Ortho-II ordered YBCO.  Here a superstructure peak at [0.5~0~0] corresponding to the Ortho-II oxygen superstructure is clearly seen.  In Fig.~\ref{fig1} a), the scattering intensity at the [0.5~0~0] Ortho-II superstructure Bragg reflection is shown as a function of incident photon energy through the Cu $L_3$ ($\sim 930$ eV) and $L_2$ ($\sim 950$ eV) x-ray absorption edges for both $\sigma$ and $\pi$ incident polarization.  At the absorption edges, the scattering intensity increases dramatically.\cite{Feng04}  In Figs.~\ref{fig1} b) and c) the intensity is shown as a function of $H$ and $K$ about the [0.5~0~0] peak.  As discussed in references \onlinecite{Zimmermann03} and \onlinecite{Schleger95}, the Ortho-II superstructure peak fits well to
\begin{equation}
I(\vec{q}) = \frac{A}{\left(1+\left(\frac{\Delta q_h}{\Gamma_h}\right)^2+\left(\frac{\Delta q_k}{\Gamma_k}\right)^2+\left(\frac{\Delta q_l}{\Gamma_l}\right)^2\right)^y} + B,
\label{eq:fit}
\end{equation}
where $\Delta \vec{q}$ is the reduced momentum transfer (= $\vec{q}$ - [0.5~0~0]) and $\Gamma_i$ is inverse correlation length, related to the correlation length $\xi_i$ by $\Gamma_h = a/2\pi\xi_h$ (and likewise for $k$ and $l$).   The finite experimental momentum resolution, which is $< 5\%$ of $\Gamma$, is neglected.  
Our measurements are fit well by equation~\ref{eq:fit} giving correlation lengths of $\xi_h = 31$ \AA,  $\xi_k$ =  93 \AA ~and  $\xi_l$ = 12 \AA ~and $y \simeq 1.2 - 1.5$.  These are comparable to previous hard x-ray results, with the value of $y$ consistent with 2D ordering of the oxygen into finite size domains that have sharp boundaries.\cite{Zimmermann03,Schleger95}   

In our measurements, the additional constant background, $B$, is dominated by sample fluorescence, which tracks the x-ray absorption and thus changes as a function of incident photon energy, as seen clearly in Figs.~\ref{fig1} b) and c).  The shape and width of the [0.5~0~0] Bragg reflection does not change appreciably as a function of incident photon energy.  As such, the peak intensity (shown in Fig.~\ref{fig1} a) has the same energy dependence as the integrated peak intensity (the fluorescent XAS background has been subtracted off the curves shown in Fig.~\ref{fig1} a).  

In principle, the energy dependence of the RSXS as a function of polarization contains detailed information about which electronic states are contributing to the superstructure being probed.  Gleaning this information from the data requires some additional knowledge about the structure and electronic structure of the material.  In this case, the Ortho-II ordering structure factor is well understood and x-ray absorption measurements can be used to calculate the atomic scattering form factor.  

Accounting for the Ortho-II structure, the scattering intensity for the [0.5~0~0] Bragg reflection is given by 
\begin{equation}
I[0.5 ~0 ~0](\omega,\vec{\epsilon}) \propto \left|f_{Cu1f}(\omega,\vec{ \epsilon }) +f_O-f_{Cu1e}(\omega,\vec{\epsilon})\right|^2
\label{eq:Iscatt}
\end{equation}
where  $f_{Cu1f}(\omega,\vec{ \epsilon })$, $f_{Cu1e}(\omega,\vec{ \epsilon }) $ and $f_O$ are the atomic scattering form factors for the Cu in the full chain, Cu in the empty chains and oxygen, respectively.  Additional smaller contributions to the scattering intensity coming from displacements of Ba and Y, or from potential charge ordering induced on the Cu in the CuO$_2$ planes, are neglected (this assumption will be returned to).  Notably, the Cu1 and Cu2 cites are not subject to lattice distortions away from their ideal positions in the Ortho II phase, making Eqn.~\ref{eq:Iscatt} dependent only upon the difference $f_{Cu1f}(\omega,\vec{ \epsilon }) - f_{Cu1e}(\omega,\vec{\epsilon}$).

On resonance, the atomic scattering form factor for each atom is given by
\begin{equation}
f_n(\omega, \vec{\epsilon}) = \vec{\epsilon'} \cdot F_n(\omega)\vec{\epsilon},
\end{equation}
where $\vec{\epsilon}$ and $\vec{\epsilon'}$ are the incoming and scattered polarization, respectively.  The 3$\times$3 scattering tensor $F_n$ captures any anisotropy in the scattering form factor.  Although the outgoing polarization was not measured, the local symmetry of the Cu1e and Cu1f sites imply that only $\sigma\sigma^\prime$ or $\pi\pi^\prime$ scattering is allowed.  In this instance and for the geometry used, $f(\omega, \sigma \sigma^\prime) = f_c(\omega)$ and  $f(\omega, \pi \pi^\prime) = f_a(\omega)\cos^2\theta - f_b(\omega)\sin^2\theta$, where $f_a$, $f_b$ and $f_c$ are the diagonal components $F_n$.

On resonance, the atomic scattering form factors, $f(\omega,\vec{\epsilon})=f_1(\omega,\vec{\epsilon}) + if_2(\omega,\vec{\epsilon})$,  are complex functions with $f_2(\omega,\vec{\epsilon})$ proportional to the linear absorption co-efficient, $\mu(\omega,\vec{\epsilon})$:
\begin{equation}
f_2(\omega,\vec{\epsilon}) = -\frac{\omega m_ec}{4\pi e^2 n_v}\mu(\omega,\vec{\epsilon}),
\end{equation}
where $n_v$ is the molar volume.  $f_1(\omega,\vec{\epsilon})$ can be deduced from $f_2(\omega,\vec{\epsilon})$ using the Kramers-Kronig transformation:
\begin{equation}
f_1(\omega ) = Z + \frac{2}{\pi } \oint_0^\infty \frac{tf_2(t)}{t^2 - \omega^2} dt,
\label{eq:KK}
\end{equation}
where $Z$ is the charge on the atom.  

While it is clear that the resonant scattering intensity can in principle be calculated from the x-ray absorption, a complication arises in that the measured x-ray absorption is a sum over contributions from all atoms in the material.  As such, it is often difficult to calculate the resonant scattering intensity from the x-ray absorption without some means to isolate the contribution to the total x-ray absorption from individual atoms.  In Ortho-II ordered YBCO, we make use of x-ray absorption measurements on different samples that have either entirely empty chains ($\delta = 0$) or entirely full chains ($\delta = 1$) in order to deduce $f_{Cu1f}$ and $f_{Cu1e}$.

\begin{figure}[t]
\centering
\resizebox{\columnwidth}{!}{\includegraphics{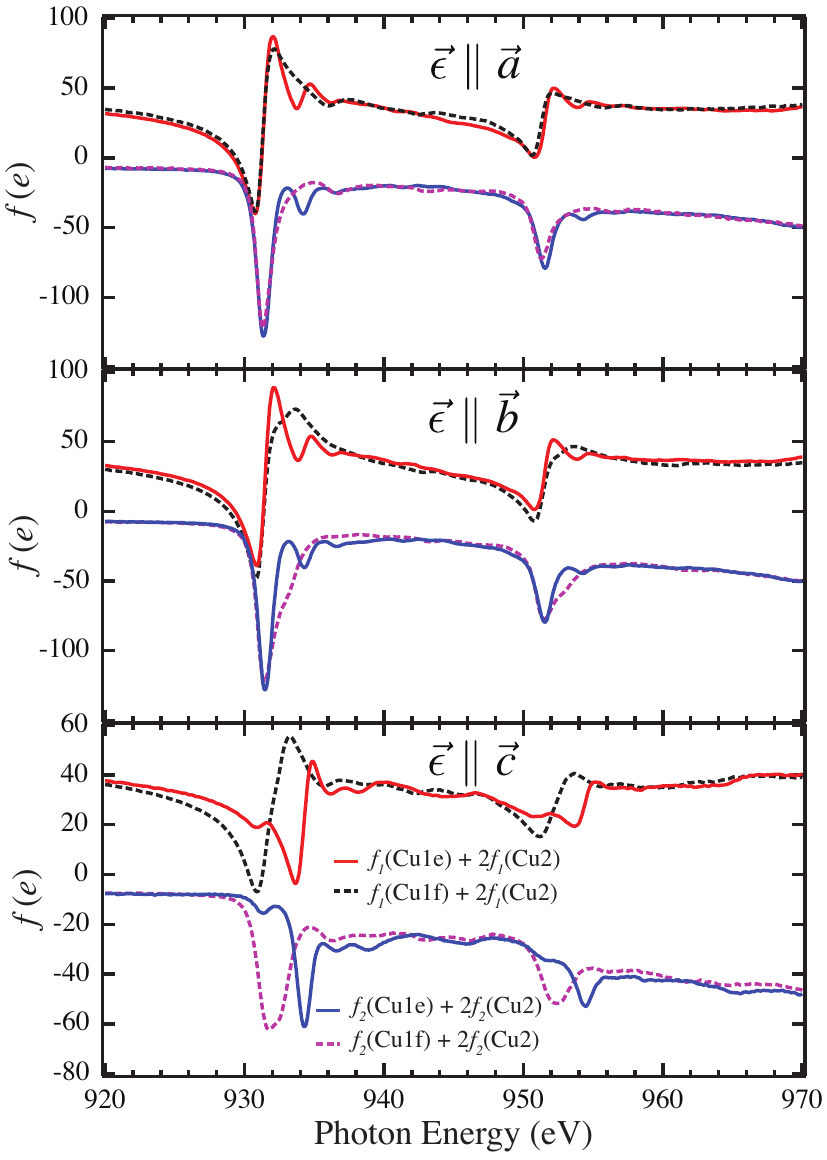}}
\caption{(colour online) The real ($f_1$) and imaginary ($f_2$) components of the scattering form factor calculated from the x-ray absorption of undoped and fully doped YBCO for $\vec{\epsilon} \parallel \vec{a}$, $\vec{\epsilon} \parallel \vec{b}$ and $\vec{\epsilon} \parallel \vec{c}$.  $f$ is determined for Cu1e and Cu1f from undoped ($\delta$ = 0) and fully doped ($\delta$ = 1) YBCO, respectively.  Since the XAS is the sum of contributions from both the chain and plane sites, the calculated scattering form factors are also the sum of Cu1 and Cu2 sites.}
\label{fig_f}
\end{figure}

In Fig.~\ref{fig_f}, $f_1$ and $f_2$ are shown for Cu1e and Cu1f for $\vec{\epsilon} \parallel \vec{a}, \vec{b}$ and $\vec{c}$.   Since the XAS is the sum of contributions from both the chain and plane sites, the calculated scattering form factors are also the sum of Cu1 and Cu2 sites.  However, in the calculation of the scattering intensity using Eqn.~\ref{eq:Iscatt}, the contributions from Cu2 sites will cancel out.  To give $f$ in units of $e$/atom, the XAS is normalized to the values of $f_2$ determined from ref.~\onlinecite{Henke93} taken below (915 eV; $f_2$ = -2.14) and above (970 eV; $f_2$ = -16.83) the Cu $L_{2,3}$ edges.  $f_O(930$ eV) = 8.1 - 2.1$i$ was also determined ref.~\onlinecite{Henke93}.   These values of $f_2$ are determined from tabulated data that does not properly capture near-edge structure.  However, sufficiently far from an absorption edge, these values are in reasonable agreement with experiment.\cite{Henke93}

For the Kramers-Kronig transform, $f$ has to be extrapolated to $\omega = 0$ and $\infty$.  This is done by setting $f_2$ to a constant equal to $f_2(915\text{ eV})$ below 915 eV and to $f_2(975\text{ eV})[\omega/975\text{ eV}]^2$ above 975 eV.  Errors in this procedure have a tendency to introduce energy independent offsets in the value of $f$.  However, because constant offsets in $f$ occur for both $f_{Cu1e}$ and $f_{Cu1f}$ and ultimately cancel in Eqn.~\ref{eq:Iscatt}, the calculated scattering intensity between 920 eV and 960 eV is only weakly sensitive to different extrapolation methods.

\begin{figure}[th]
\centering
\resizebox{\columnwidth}{!}{\includegraphics{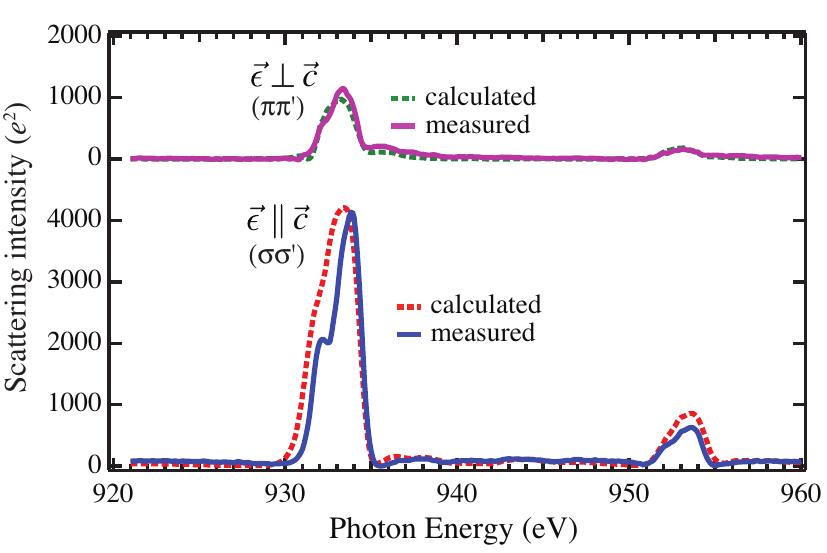}}
\caption{(colour online) A comparison of the measured and calculated scattering intensities as a function of incident photon energy through the Cu $L$ edges for both $\vec{\epsilon} \parallel \vec{c}$ ($\sigma\sigma^\prime$) and $\vec{\epsilon} \perp \vec{c}$ ($\pi\pi^\prime$) scattering geometries.  The measured $\vec{\epsilon} \parallel \vec{c}$ spectrum was scaled in intensity to roughly match the calculated spectrum, which is expressed in units of $e^2$ (following eqn.~\ref{eq:Iscatt}).   The $\vec{\epsilon} \perp \vec{c}$ spectrum was scaled by the same amount as the $\vec{\epsilon} \parallel \vec{c}$ spectrum and offset for clarity.  This scaling parameter is the only free parameter.  The calculation agrees well with the data, capturing the relative intensities of the $L_3$ and $L_2$ edges, as well as the relative intensity of the $\vec{\epsilon} \parallel \vec{c}$ and $\vec{\epsilon} \perp \vec{c}$ scattering geometries.}
\label{fig4}
\end{figure}

The calculated resonant x-ray scattering intensity as a function of energy is shown in Fig.~\ref{fig4} along with the measured spectra for $\sigma\sigma^\prime$ ($\vec{\epsilon} \parallel \vec{c}$) and $\pi\pi^\prime$ ($\vec{\epsilon} \perp \vec{c}$) scattering geometries.  The agreement between the calculated and measured spectra is very good, with accurate reproducibility of the relative intensities of the $\vec{\epsilon} \parallel \vec{c}$ and $\vec{\epsilon} \perp \vec{c}$ spectra, as well as the relative intensity of the $L_3$ and $L_2$ edges.  This agreement provides strong confirmation of our approach of using XAS measurements to interpret the resonant scattering. 

The main difference between the experiment and calculations seems to be an extra broadening in the calculated scattering intensity that is not apparent in the measurement.  This additional broadening was also observed at the Cu $L_3$ edge in La$_{2-x}$Ba$_x$CuO$_4$ and La$_{1.8-x}$Eu$_{0.2}$Sr$_x$CuO$_4$,\cite{Abbamonte05,Fink09} suggesting it may be a generic feature of resonant elastic x-ray scattering.  This may occur if a different core-hole lifetime broadening enters into XAS and RSXS.  In particular, the elastic scattering process, which requires the energy of the incident and scattered photons to be the same, may act in a similar fashion to using high resolution fluorescence spectroscopy to eliminate core-hole broadening in XAS spectra.\cite{Hamalainen91}

The good agreement between experiment and calculation also indicates that the scattering intensity is almost entirely the result of charge ordering in the chains, since we have neglected charge ordering in the planes. This conclusion is in contrast to the results of Feng et al., who argued for a sizeable in-plane modulation based on the polarization dependence of the Cu $L$ edge RSXS lineshape.\cite{Feng04}  However, contrary to the analysis of Feng {\it et al.}, our detailed analysis of the resonant scattering lineshape indicates that the measurements do not have sufficient sensitivity to the CuO$_2$ planes to address the issue of in-plane ordering.   
\begin{figure}[th]
\centering
\resizebox{\columnwidth}{!}{\includegraphics{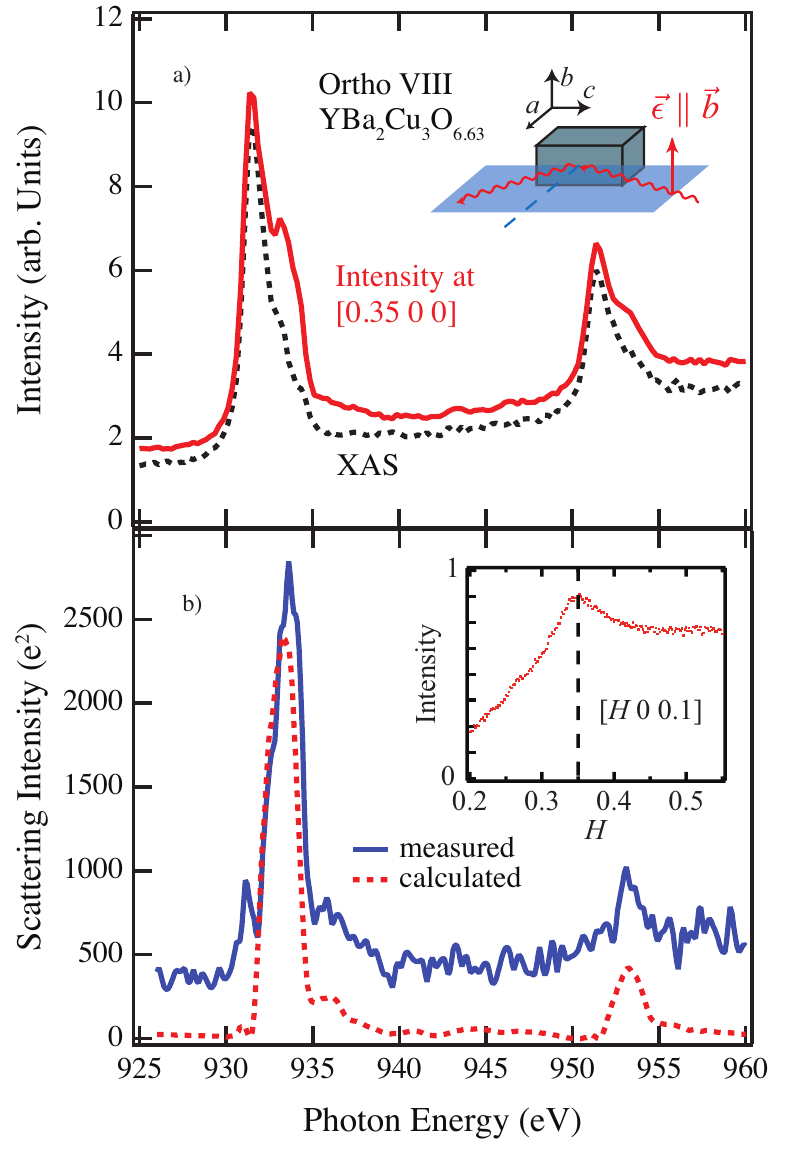}}
\caption{(colour online) a) The measured intensity vs photon energy at the [0.35 0 0] superlattice peak of Ortho-VIII ordered YBCO.  A large background (black) from x-ray fluorescence measured away from the superlattice peak contributes to the measured intensity at [0.35 0 0].  b)  Subtracting out the x-ray fluorescence reveals the measured scattering intensity vs photon energy through the Cu $L$ edge.  Apart from a constant offset, the measured scattering intensity (blue) agrees well with the calculated intensity (red).}
\label{fig5}
\end{figure}

Similar to the chain contribution to the resonant x-ray scattering, contributions to the RSXS cross-section from the CuO$_2$ planes can be determined from a Kramers-Kronig transform of the x-ray absorption.  Similar to the hole modulation of the Cu1 in the chain layer, for a hole modulation in the Cu2 sites, the structure factor is such that the scattering amplitude should depend on the difference in $f(\omega)$ at the different Cu2 sites.  In contrast, if the Cu2 were subject to a lattice displacement from their ideal lattice positions, the structure factor is such that scattering intensity will depend on the overall magnitude of  $f(\omega)$ at the Cu2 sites, producing a peak in the scattering roughly where the Cu $L$ XAS is peaked.  This latter case is consistent with the scattering intensity at the Cu $L$ edge observed at the CDW superlattice in La$_{2-x}$Ba$_x$CuO$_4$ and La$_{1.8-x}$Eu$_{0.2}$Sr$_x$CuO$_4$.\cite{Abbamonte05,Fink09}  However, because of the crystal symmetry about the Cu2 site in Ortho-II YBCO, one would not expect a lattice displacement of Cu2 due to the Ortho-II order and the scattering amplitude should depend on the difference in $f(\omega)$ at the different Cu2 sites.

As noted above, the doping of charge carriers in the CuO$_2$ planes is evidenced in the Cu $L$ edge XAS by a shoulder that arises off the main peak at $\sim 931.6$ eV for $\vec{\epsilon} \parallel \vec{a}$ or $\vec{b}$ (see fig.~\ref{fig2} and fig.~\ref{fig_f}).  Comparison with XAS measurement on LSCO, which has only a single in-plane Cu site, indicates that this shoulder feature exhibits the largest changes with doping (a weaker feature with $\vec{\epsilon} \parallel \vec{c}$ also shows some doping dependence).\cite{Chen91}   
From fig.\ref{fig_f} (top panel) we can quantify the magnitude of the difference in the scattering form factor for the Cu2 sites at 931.6 eV.  Around 931.6 eV, the maximum difference between $f_2(\delta = 1)$ and  $f_2(\delta = 0)$ is 13.5$e$.  This difference corresponds to a difference of $\sim 0.2$ electrons per Cu in the plane (the difference in the in-plane hole doping between undoped and fully doped YBCO).  However, any expected modulation of charge in the CuO$_2$ planes due to the Ortho-II order will be considerably less than $\sim 0.2$ electrons per Cu.   For the purpose of producing an estimate of the scattering amplitude we will assume a modulation of 0.02 electrons per Cu in the CuO$_2$ planes.  In addition, for $\pi\pi^\prime$ scattering with $\theta \simeq 60^\circ$, the scattering amplitude from the CuO$_2$ planes is reduced by $(\sin^2\theta -\cos^2\theta)^2 \simeq 1/4$.  Thus, the scattering amplitude in units of $f(\omega)^2$ from the planes can be estimated to be $\sim 1/4\times(0.02/0.2 \times 13.5 e)^2 \simeq 0.5 e^2$.  This contribution to the scattering would be very difficult to detect when compared with the scattering from the chains, which has a magnitude of $\sim 1000 e^2$ around 931.6 eV.  

Finally, in Fig.~\ref{fig5} we present the scattering intensity as a function of photon energy for the [0.35, 0 0] superlattice peak of Ortho-VIII.  The intensity of the superlattice peak is significantly weaker in Ortho-VIII relative to Ortho-II ordered YBCO.  The measured intensity as a function of photon energy at [0.35 0 0] is largely due to x-ray fluorescence rather than scattering.  Subtracting the fluorescence (measured at a $Q$ vector away from the superlattice peak) from the data, however, yields the scattering intensity as a function of photon energy (shown in Fig.~\ref{fig5} b).  Like the Ortho-II measurements, this result can also be compared to the calculated scattering intensity (here the geometry is different and $f(\omega) = f_b(\omega)$). Aside from an energy independent offset in the measurement, the measured and calculated scattering are in good agreement, confirming that the [0.35 0 $L$] peak in Ortho-VIII is also from the chain layer and is not due to CDW order in the CuO$_2$ planes.

In conclusion, we report a null result in our search for static CDW order in the CuO$_2$ planes of underdoped, oxygen-ordered YBCO.  This places restrictions on the parameter space in which static CDW order may exist in YBCO.  In addition, in Ortho-II ordered YBCO we measured the polarization and energy dependence of the [0.5~0~0] scattering intensity through the Cu $L$ edges.  These measurements are in excellent agreement with calculations of scattering intensity using polarization and energy dependent atomic scattering forms factors deduced from a simple Kramers-Kronig transform of the measured x-ray absorption in undoped and fully doped YBCO, which have empty and full chains respectively.  This comparison between calculation and experiment confirms the validity of the approach to the interpretation of the resonant x-ray scattering lineshapes.  In addition, these results confirm our understanding of the hole ordering in the chains of oxygen-ordered YBCO.

\begin{acknowledgments}
This work was supported by the Canadian Institute for Advanced Research, the British Columbia Synchrotron Institute, a Canada Research Chair (GAS), the Canada Foundation for Innovation, and the Natural Sciences and Engineering Research Council of Canada.  J.G. gratefully acknowledges the financial support through the DFG.
\end{acknowledgments}


\end{document}